\documentclass[aps,11pt]{revtex4}
\usepackage{dcolumn}
\usepackage{graphicx}
\usepackage{amsmath}
\usepackage{amsfonts}
\usepackage{amssymb}
\usepackage{psfrag}
\usepackage{wrapfig}
\usepackage{subfigure}
\usepackage{makeidx}
\usepackage{bm}
\usepackage{epsf}
\usepackage{hyperref}
\usepackage{color}
\usepackage{cases}

\makeatletter

\makeatletter
\newcommand{\eq}[1]{Eq.(\ref{#1})}
\newcommand{\ud}{\,\mathrm{d}\,}

\newcommand{\Rmnum}[1]{\uppercase\expandafter{\romannumeral #1}}

\begin{document} 

\title{Quasinormal Modes of a black hole surrounded by a fluid of strings in Rastall gravity}

\author{Ming Zhang$^{1}$\footnote{e-mail: zhangming@xaau.edu.cn; mingzhang0807@126.com} and Rui-Hong Yue$^{2}$\footnote{e-mail: rhyue@yzu.edu.cn}, De-Cheng Zou$^{3}$\footnote{e-mail: dczou@jxnu.edu.cn}}

\address{$^{1}$Faculty of Science, Xi'an Aeronautical University, Xi'an 710077 China\\
$^{2}$Center for Gravitation and Cosmology, College of Physical Science and Technology, Yangzhou University, Yangzhou 225009, China\\
$^{3}$College of Physics and Communication Electronics, Jiangxi Normal University, Nanchang 330022, China}

\date{\today}

\begin{abstract}
\indent

In this paper, we explore the quasinormal modes (QNMs) of a black hole surrounded by a fluid of strings within the framework of Rastall gravity. We analyze the behavior of scalar, electromagnetic, and gravitational perturbations, focusing on the influence of the black hole charge $Q$ and angular momentum $l$ on the quasinormal frequencies. Our numerical results reveal a significant dependence on the parameter $\varepsilon$. These trends are consistent across different types of perturbations, emphasizing the relationship between black hole parameters and QNMs behavior.
\end{abstract}


\maketitle

\section{Introduction}
\label{intro}

General relativity has withstood extensive scrutiny since its predictions were first confirmed in 1919, particularly through phenomena like the deflection of light by the Sun. Recent evidence from gravitational wave observations by LIGO and Virgo further supports its robustness\cite{Will:2014kxa,Hulse:1974eb,Damour:1991rd}. However, general relativity does not account for dark matter and energy or the cosmic acceleration \cite{SupernovaSearchTeam:1998fmf,SupernovaCosmologyProject:1998vns}, and it is not renormalizable\cite{Stelle:1976gc}.
These limitations motivate the exploration of modified theories, one of which is Rastall gravity\cite{Rastall:1972swe}. Rastall proposed a modification to the energy-momentum conservation law, suggesting that its divergence could be proportional to the gradient of the Ricci scalar\cite{Rastall:1972swe}. This leads to a non-minimal coupling of matter with spacetime geometry, recovering standard conservation in flat spacetime. Numerous studies suggest Rastall gravity is a viable alternative, with applications in astrophysics and cosmology\cite{Oliveira:2015,Pereira:2010,Calogero:2011,Calogero:2013,Moradpour:2016,Al-Rawaf:1996a,Al-Rawaf:1996b,Moradpour:2017}. Notably, solutions involving black holes surrounded by a cloud of strings have been explored, revealing interesting properties such as critical rotation parameters for extremal black holes\cite{Kumar:2018} and stability of certain gravastar models\cite{Ghosh:2021}. The concept of a string cloud as a source of gravitational fields dates back to Letelier, who developed solutions to the Einstein equations for various symmetries\cite{Letelier:1979,Barbosa:2016}. This work laid the groundwork for later studies that incorporated pressure effects\cite{Letelier:1981} and anisotropic fluids \cite{Soleng:1995,Dymnikova:1992,Soleng:1994}. The potential of Rastall gravity to modify spacetime properties makes it a compelling area for further investigation, especially in conjunction with models involving strings.

Quasinormal modes (QNMs) are essential characteristics of dissipative systems, particularly in black holes, where anything that crosses the event horizon cannot escape. QNMs dominate the ringdown phase of gravitational waves from binary black hole mergers\cite{Berti:2007}. Unlike normal modes, the eigenfunctions of QNMs do not form a complete set and are not normalizable\cite{Nollert:1999}. QNMs exhibit complex frequencies, with real parts representing vibration frequencies and imaginary parts indicating decay time scales. Studying QNMs is crucial for inferring the mass and angular momentum of black holes, as well as testing the no-hair theorem\cite{Berti:2006,Berti:2007b,Isi:2019}. In horizonless compact objects, QNMs may reveal echoes in the ringdown signal, providing evidence for their existence \cite{Cardoso:2017,Cardoso:2016,Cardoso:2019}. Additionally, QNMs can constrain modified gravity theories\cite{Wang:2004,Blazquez-Salcedo:2016,Franciolini:2019,Aragon:2021,Liu:2021,Karakasis:2022,Cano:2022,Gonzalez:2022upu,Zhao:2022gxl} and have been found to be unstable under small potential perturbations\cite{Jaramillo:2021,Cheung:2022}. They also help assess the stability of the background spacetime under perturbations\cite{Ishibashi:2003,Chowdhury:2022zqg}.

In this work, we focus on the black hole surrounded by a fluid of strings in Rastall gravity that presented in Ref.\cite{Bezerra:2022srj}. We investigates the QNMs of the Rastall black hole under three distinct types of perturbations of scalar, electromagnetic, and gravitational fields. 
We analyze the effects of the black hole parameters on the quasinormal frequencies, revealing how these parameters influence the stability and decay properties of perturbations. This paper is constructed as follows: we first review the solutions of a black hole surrounded by a fluid of strings in Rastall gravity in Section \ref{sec2}. Then Section \ref{sec3} and \ref{sec4} give the pertubation equations of three fields and detail our numerical methods for calculating the QNMs. Section \ref{sec5} presents the figures of QNMs and discusses our results.  Finally, we provide concluding remarks in the sixth section.

\section{Solution}
\label{sec2}

Considering the modification of theory of general relativity made by Peter Rastall\cite{Rastall:1972swe}
\begin{eqnarray}
  T^{\nu}_{ \mu;\nu}=\beta R_{,\nu},
\end{eqnarray}
where $\beta$ is a constant and $R$ is the Ricci scalar. The field equation reads as
\begin{eqnarray}
  R^\nu_\mu - \frac{1}{2} \delta^\nu_\mu R = k (T^\nu_\mu - \beta \delta^\nu_\mu R) \label{eom}.
\end{eqnarray}
In the limit as $\lambda \to 0$, $k = 8 \pi G_N$, where $G_N$ represents the Newtonian gravitational constant, and the field equations reduce to GR field equations.

In four-dimensional spacetime, the energy-momentum
tensor $T_{\mu\nu}$ of perfect fluid reads as\cite{Bezerra:2022srj}
\begin{eqnarray}
  T_{\mu\nu} = \left( q + \rho \sqrt{-h} \right) \frac{\Sigma^{\mu\lambda} \Sigma_{\lambda}^{\nu}} {(-h)} + q g^{\mu\nu}
\end{eqnarray}
the parameters $q$ and $\rho$ represent the pressure and density of the fluid of strings, respectively.

Then the solution of black hole surrounded by a fluid of strings in four-dimensional Rastall gravity is\cite{Bezerra:2022srj}
\begin{eqnarray}
  ds^2 = -f(r) \, dt^2 + \frac{1}{f(r)} \, dr^2 + r^2 \,( d\theta^2 + \sin^2 \theta \, d\phi^2),
\end{eqnarray}
and
\begin{eqnarray}
  f(r) = 1 &-& \frac{2M}{r} + \frac{Q^2}{r^2} \nonumber\\
 &-& \begin{cases} 
   \varepsilon L r^{-1} \log(\lambda r) & \text{for } A = 2, \\
   \varepsilon A (A - 2)^{-1} \left( \frac{L}{r} \right)^{2/A} & \text{for } A \neq 2,
   \end{cases}
\end{eqnarray}
where $\lambda$ and $L$ is a positive constant, $Q$ represents the electric charge. $\varepsilon=\pm 1$ gives the sign of the energy density of the fluid of strings and $A$ is defined as
\begin{eqnarray}
  A = \frac{\alpha - 2k\beta(1 + \alpha)}{1 - 2k\beta(1 + \alpha)}
\end{eqnarray}
where $\alpha$ is a dimensionless constant.

\section{Perturbation Equations}
\label{sec3}

In this section, we examine the linear perturbation equations for the scalar, electromagnetic, and gravitational fields in the background of a black hole surrounded by a fluid of strings in Rastall gravity. For clarity, we consider each type of perturbed field independently, assuming that perturbing one field does not affect the background of the other two fields.

\subsection{Scalar perturbations}

We consider the evolution of a free massless scalar field, the equation for scalar in the background of this black hole is given by 
\begin{eqnarray}
  \frac{1}{\sqrt{-g}} \, \partial_\mu \left( \sqrt{-g} \, g^{\mu\nu} \partial_\nu \varphi\right)=0 \label{eqkg}
\end{eqnarray}
To do the separation of variables we introduce the radial function $\varphi_s(r)$ and spherical harmonics $Y_{lm}(\theta,\phi)$ as
\begin{eqnarray}
  \varphi(t,r,\theta,\phi)=\sum_{lm}\frac{\varphi_s(r)}{r}Y_{lm}(\theta,\phi)e^{-i\omega t} \label{eqphis}
\end{eqnarray}
By substituting Eq.{\ref{eqphis}} to Eq.{\ref{eqkg}}, the radial perturbed equation can be written as the Schr\"{o}dinger wavelike form
\begin{eqnarray}
  \frac{\ud^2 \varphi_s(r_*)}{\ud r^2_{*}}+\left[\omega^2-V_s(r)\right]\varphi_s(r_*)=0 \label{pe1}
\end{eqnarray}
where the ``tortoise coordinate'' $r_{*}$ is defined as
\begin{eqnarray}
  \ud r_* \equiv \frac{\ud r}{f(r)}.
\end{eqnarray}
The effective potentials of scalar field is given by
\begin{eqnarray}
  V_s(r)=f(r)\left(\frac{l(l+1)}{r^2}+\frac{f'(r)}{r} \right).
\end{eqnarray}
where $l=0,1,2,...$ are the multipole numbers.

\subsection{Electromagnetic perturbations}

The equation for electromagnetic in the background of this black hole can be written as
\begin{eqnarray}
  \frac{1}{\sqrt{-g}} \, \partial_\mu \left( \sqrt{-g} \, g^{\sigma\mu}g^{\rho\nu} F_{\rho\sigma}\right)=0 \label{eqmax}
\end{eqnarray}
where $F_{\rho\sigma}=\partial_{\rho}A_{\sigma}-\partial_{\sigma}A_{\rho}$ and $A_{\mu}$ is the field-strength tensor of the perturbed electromagnetic field which can be decomposed as
\begin{eqnarray}
  A_\mu(t, r, \theta, \phi) = \sum_{l,m} e^{-i\omega t}
\begin{bmatrix}
0 \\
0 \\
\frac{\varphi_e(r)} {\sin \theta} \frac{\partial Y_{l,m}}{\partial \phi} \\
-\varphi_e(r) \sin \theta \frac{\partial Y_{l,m}}{\partial \theta}
\end{bmatrix}
+ \sum_{l,m} e^{-i\omega t}
\begin{bmatrix}
h_1(r) Y_{l,m} \\
h_2(r) Y_{l,m} \\
h_3(r) \frac{\partial Y_{l,m}}{\partial \theta} \\
h_3(r) \frac{\partial Y_{l,m}}{\partial \phi}
\end{bmatrix} \label{eqA}
\end{eqnarray}
Due to the spherical symmetry of the background metric, the perturbation equations separate polar and axial contributions. Furthermore, the axial and polar parts have equal contributions to the final result\cite{Wheeler:1973,Ruffini:1973}. Therefore, we only need to focus on the axial part.

By substituting \eq{eqA} to \eq{eqmax}, the perturbation equation for the radial part $\varphi_e(r)$ can be obtained as the Schr\"{o}dinger wavelike form by using the tortoise coordinate $r_*$
\begin{eqnarray}
  \frac{\ud^2 \varphi_e(r_*)}{\ud r^2_{*}}+\left[\omega^2-V_e(r)\right]\varphi_e(r_*)=0 \label{pe2}
\end{eqnarray}
where the effective potentials of electromagnetic field is given by
\begin{eqnarray}
  V_e(r)=\frac{f(r)l(l+1)}{r^2}.
\end{eqnarray}
The EM modes exist for $l \geq 1$.

\subsection{Gravitational perturbations}

We consider the first-order gravitational perturbed fields around the background quantities
\begin{eqnarray}
  \bar{g}_{\mu\nu}=g_{\mu\nu}+h_{\mu\nu}
\end{eqnarray}
where $h_{\mu\nu}$ is the perturbation. The separation of perturbations for the gravitational field is more complicated. This decomposition divides the tensor-vector perturbations into ``axial (A)'' modes, which gain a factor of $(-1)^{l+1}$ under parity inversion, and ``polar (P)'' modes, which gain a factor of $(-1)^{l}$.

Under the Regge–Wheeler gauge\cite{Regge:1957}, we expand the metric perturbations using tensor spherical harmonics. For the axial part (odd parity), which involves two modes $h_0$ and $h_1$, the perturbed metric is expressed as
\begin{eqnarray}
  h_{\mu\nu} = \sum_{l,m} e^{-i\omega t}
\begin{bmatrix}
0 & 0 & -\frac{h_0(r)\partial_\varphi Y_{l,m}}{\sin \theta} & h_0(r)\sin \theta \partial_\theta Y_{l,m} \\
0 & 0 & -\frac{h_1(r)\partial_\varphi Y_{l,m}}{\sin \theta} & h_1(r)\sin \theta \partial_\theta Y_{l,m} \\
-\frac{h_0(r)\partial_\varphi Y_{l,m}}{\sin \theta} & -\frac{h_1(r)\partial_\varphi Y_{l,m}}{\sin \theta} & 0 & 0 \\
h_0(r)\sin \theta \partial_\theta Y_{l,m} & h_1(r)\sin \theta \partial_\theta Y_{l,m} & 0 & 0
\end{bmatrix} \label{eqodd}
\end{eqnarray}

For polar perturbations (even parity) with four modes $(H_0, H_1, H_2, K)$, we have
\begin{eqnarray}
  h_{\mu\nu} = \sum_{l,m} e^{-i\omega t}
\begin{bmatrix}
H_0(r)f(r) & H_1(r) & 0 & 0 \\
H_1(r) & \frac{H_2(r)}{f(r)} & 0 & 0 \\
0 & 0 & r^2 K(r) & 0 \\
0 & 0 & 0 & r^2\sin^2 \theta  K(r)
\end{bmatrix}
Y_{l,m}.
\end{eqnarray}
Note that we can chose $m=0$ in the later calculation for simplicity, as the perturbation equations are independent of the value of $m$ \cite{Regge:1957}.

For the Schwarzschild black hole, ref.\cite{Chandrasekhar:1983} shows that the odd and even parities share the same QNM spectrum, though this may not be true for other black holes. In this paper, we simplify our analysis by focusing on the odd parity. By substituting the decomposition into Einstein's equations and performing some algebraic manipulations, the perturbation equations for odd parity can be consolidated into a single equation for the variable $\varphi_{g}$, which can be written as
\begin{eqnarray}
  \varphi_{g}=\frac{f(r)}{r}h_1(r) \label{phig}
\end{eqnarray}
By substituting \eq{phig} and \eq{eqodd} into the equations of motion \eq{eom}, after some algebra calculations and simplify we can obtain the following master perturbation equations as the Schr\"{o}dinger wavelike form by using the tortoise coordinate $r_*$
\begin{eqnarray}
  \frac{\ud^2 \varphi_g(r_*)}{\ud r^2_{*}}+\left[\omega^2-V_g(r)\right]\varphi_g(r_*)=0 \label{pe3}
\end{eqnarray}
where the effective potentials of gravitational field is given by
\begin{eqnarray}
  V_g(r)=f(r)\left(\frac{l(l+1)}{r^2}-\frac{f'(r)}{r}+f''(r) \right).
\end{eqnarray}
The gravitational modes exist for $l \geq 2$.

\section{Quasinormal Modes}
\label{sec4}

We are going to employ the improved asymptotic iteration method (AIM)\cite{Cho:2009cj} to solve the perturbation equations \eq{pe1}, \eq{pe2} and \eq{pe3} numerically. To achieve this, we take scalar perturbation \eq{pe1} for example and rewrite it in terms of $u=1-r_+/r$
\begin{eqnarray}
  \varphi_s''(u)+\left(\frac{f'(u)}{f(u)}-\frac{2}{1-u} \right)\varphi_s'(u)+\left[ \frac{r_+^2\omega^2}{(1-u)^4 f(u)^2}-\frac{l(l+1)}{(1-u)^2 f(u)}-\frac{f'(u)}{(1-u)f(u)} \right]\varphi_s(u)=0. \label{equ1}
\end{eqnarray}
such that, the range of $u$ satisfies $0\leqslant u<1$.
At the black hole horizon, the boundary conditions are pure ingoing waves
$(\varphi_s\sim e^{-i\omega r_*},~ r_*\to -\infty)$,
and pure outgoing waves
$(\varphi_s\sim e^{i\omega r_*},~ r_*\to +\infty)$,
at the spatial infinity.

To propose an ansatz for \eq{equ1}, we will examine the behavior of the function $\varphi_s(u)$ at horizon $(u=0)$ and at the boundary $u=1$. Near the horizon $(u=0)$, we have $f(0)\approx u f'(0)$. Thus, \eq{equ1} reduces to
\begin{eqnarray}
  \varphi_s''(u)+\frac{1}{u}\varphi_s'(u)+\frac{r_+^2\omega^2}{u^2 f'(0)^2}\varphi_s(u)=0.
\end{eqnarray}
The solution of this equation can be written as
\begin{eqnarray}
  \varphi_s(u\to 0)\sim C_1 u^{-\xi}+C_2 u^{\xi},~ \xi=\frac{ir_+\omega}{f'(0)},
\end{eqnarray}
where we have to set $C_2=0$ in order to respect the ingoing condition at the black hole horizon.

At infinity $(u=1)$, the asymptotic form of \eq{equ1} can be written as
\begin{eqnarray}
  \varphi_s''(u)-\frac{2}{1-u}\varphi_s'(u)+\frac{r_+^2\omega^2}{(1-u)^4}\varphi_s(u)=0.
\end{eqnarray}
which has the solution
\begin{eqnarray}
  \varphi_s(u\to 1)\sim D_1 e^{-\zeta}+D_2 e^{\zeta},~ \zeta=\frac{ir_+\omega}{1-u},
\end{eqnarray}
In order to impose the outgoing boundary condition, we should set $D_1=0$.

Now, using the above solutions at horizon and infinity, we can define the general ansatz for \eq{equ1} as
\begin{eqnarray}
  \varphi_s(u)=u^{-\xi}e^{\zeta}\chi(u) \label{eqschi}.
\end{eqnarray}
Substitute \eq{eqschi} to \eq{equ1}, we have
\begin{eqnarray}
  \chi''=\lambda_0(u)\chi'+s_0(u)\chi \label{eqschieq},
\end{eqnarray}
where 
\begin{eqnarray}
  \lambda_0(u)=\frac{2 i r_+ \omega }{u f'(0)}-\frac{f'(u)}{f(u)}-\frac{2 \left(i r_+ \omega +u-1\right)}{(1-u)^2}
\end{eqnarray}
and
\begin{eqnarray}
  s_0(u)&=&\frac{1}{(1-u)^4 u^2 f(u)^2 f'(0)^2}\Big[ -r_+^2 u^2 \omega ^2 f'(0)^2 \nonumber\\
  &+&u (u-1)^2 f(u) f'(0) \left(l (l+1) u f'(0)+f'(u) \left(i r_+ (u-1)^2 \omega -u f'(0) \left(i r_+ \omega +u-1\right)\right)\right) \nonumber\\
  &+&r_+ \omega  f(u)^2 \left(r_+ \omega  \left(-u \left(f'(0)+2\right)+u^2+1\right)^2+i (u+1) (u-1)^3 f'(0)\right) \Big]
\end{eqnarray}
Now we get the functions $\lambda_0$ and $s_0$ for the scalar perturbations and \eq{eqschieq} can be solved numerically by using the improved AIM. Following the same procedure showed above, we can also obtain the functions $\lambda_0$ and $s_0$ for the electromagnetic perturbations: 
\begin{eqnarray}
  \lambda_0(u)&=&\frac{2 i r_+ \omega }{u f'(0)}-\frac{f'(u)}{f(u)}-\frac{2 \left(i r_+ \omega +u-1\right)}{(1-u)^2}; \\
  s_0(u)&=&\frac{1}{(1-u)^4 u^2 f(u)^2 f'(0)^2}\Big[ -r_+^2 u^2 \omega ^2 f'(0)^2 \nonumber\\
  &+&(u-1)^2 u f(u) f'(0) \left(l (l+1) u f'(0)+i r_+ \omega  \left((u-1)^2-u f'(0)\right) f'(u)\right) \nonumber\\
  &+&r_+ \omega  f(u)^2 \left(r_+ \omega  \left((u-1)^2-u f'(0)\right)^2+i (u+1) (u-1)^3 f'(0)\right) \Big],
\end{eqnarray}
and 
\begin{eqnarray}
  \lambda_0(u)&=&\frac{2 i r_+ \omega }{u f'(0)}-\frac{f'(u)}{f(u)}-\frac{2 \left(i r_+ \omega +u-1\right)}{(1-u)^2}; \\
  s_0(u)&=&\frac{1}{(1-u)^4 u^2 f(u)^2 f'(0)^2}\Bigg[ -r_+^2 u^2 \omega ^2 f'(0)^2 \nonumber\\
  &+&r_+ \omega  f(u)^2 \left(r_+ \omega  \left(-u \left(f'(0)+2\right)+u^2+1\right)^2+i (u+1) (u-1)^3 f'(0)\right)  \nonumber\\
  &+&\Big[u f'(0) \left((u-1)^2 f''(u)+l (l+1)\right)+f'(u) \Big(u f'(0) \left(-i r_+ \omega +3 u-3\right)  \nonumber\\
  &+&i r_+ (u-1)^2 \omega \Big)\Big](u-1)^2 u f(u) f'(0) \Bigg],
\end{eqnarray}
for gravitational fields perturbations. In the following, we will calculate the QNM frequencies and study the influences of different model parameters on the real and imaginary parts of QNM around Rastall solutions for the scalar, electromagnetic, and gravitational fields perturbations.

\section{Numerical results}
\label{sec5}

In this section, we show the numerical results of the QNMs for the scalar, electromagnetic, and gravitational fields in the background of a black hole surrounded by a fluid of strings in Rastall gravity. 

\begin{figure}[htb]
\centering
\subfigure[$l=0$]{\label{fig:subfig:s1} 
\includegraphics[width=2.1in]{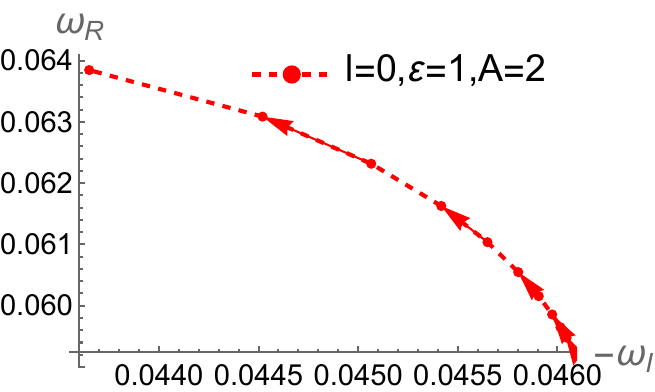}}
\hfill
\subfigure[$l=1$]{\label{fig:subfig:s2} 
\includegraphics[width=2.1in]{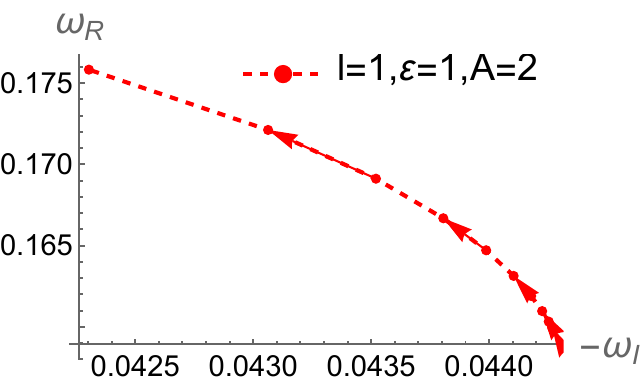}}
\hfill
\subfigure[$l=2$]{\label{fig:subfig:s3} 
\includegraphics[width=2.0in]{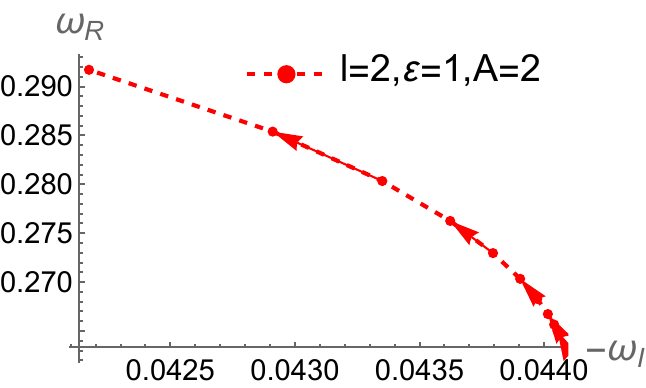}}
\hfill
\subfigure[$l=0$]{\label{fig:subfig:s11} 
\includegraphics[width=2.1in]{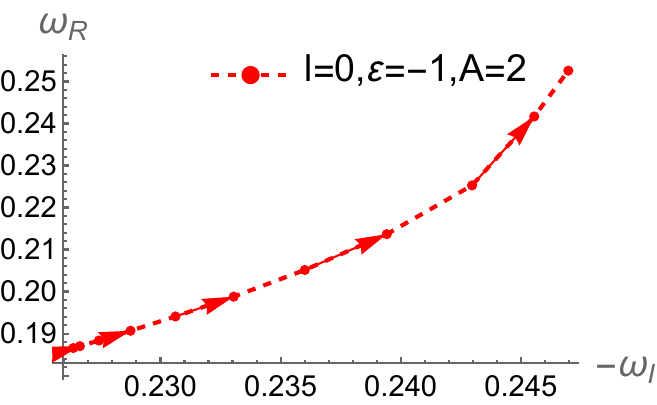}}
\hfill
\subfigure[$l=1$]{\label{fig:subfig:s21} 
\includegraphics[width=2.1in]{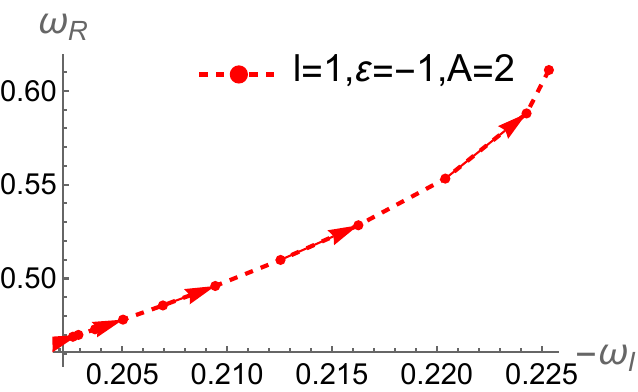}}
\hfill
\subfigure[$l=2$]{\label{fig:subfig:s31} 
\includegraphics[width=2.0in]{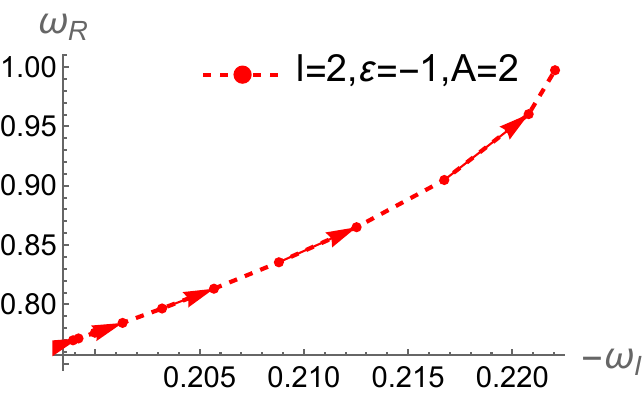}}
\hfill
\caption{The behaviors of real and imaginary parts of the \textbf{1st} branch black hole solution's QNMs under \textbf{scalar perturbations} for different $l$ with $M=1, L=1, \lambda=1$. The arrows show the direction of the increase of black hole charge $Q$ from 0 to 1.}\label{figS1}
\end{figure}

Let's begin by examining the imaginary part of the QNMs illustrated in Fig.\ref{figS1}, which show the behaviors of real and imaginary parts of the \textbf{1st} branch black hole solution's QNMs under \textbf{scalar perturbations}. When we increase the charge $Q$ of the background configuration, we notice that the absolute value of the imaginary part of the quasinormal frequency $|\omega_I|$ decreases for $\varepsilon=1$ (in Figs.\ref{fig:subfig:s1}, \ref{fig:subfig:s2} and \ref{fig:subfig:s3}) and increases for $\varepsilon=-1$ (in Figs.\ref{fig:subfig:s11}, \ref{fig:subfig:s21} and \ref{fig:subfig:s31}). Additionally, when the charge of the black hole $Q$ and other parameters of black hole are fixed, a bigger value of the angular momentum $l$ results in a smaller $|\omega_I|$. Specifically, for $\varepsilon=1$ the absolute value of the imaginary part decreases as the black hole charge increases. This indicates that the perturbations outside the black hole take a longer time to decay, meaning the perturbation remains active for an extended period before vanishing completely. From a holographic perspective, this implies that the dual system takes longer to return to equilibrium. For $\varepsilon=-1$ the behavior of $|\omega_I|$ is indeed different, which suggest the perturbations outside the black hole dissipate more rapidly as the black hole charge grows, contrary to the $\varepsilon=1$ case. This shows that the parameter $\varepsilon$ has a significant influence on the stability and decay properties of the black hole.

Let us now examine the behavior of the real part of the quasinormal frequency $|\omega_R|$ as depicted in the Fig.\ref{figS1}. For $\varepsilon=1$, as seen in panels Figs.\ref{fig:subfig:s1}, \ref{fig:subfig:s2} and \ref{fig:subfig:s3}, the real part of the quasinormal frequency $|\omega_R|$ increases as the black hole charge  $Q$ increases. The arrows pointing from right to left indicate that when $Q$ increases from $0$ to $1$, $|\omega_R|$ gradually increases. This trend holds across different values of angular momentum $l$, meaning that as the black hole becomes more charged, the scalar perturbations gain energy, leading to higher oscillation frequencies. Furthermore, for fixed $Q$, the real part $|\omega_R|$  increases with increasing angular momentum $l$, implying that perturbations with larger angular momentum have more energy, regardless of the charge $Q$. For the $\varepsilon=-1$ the behavior is similar with $\varepsilon=1$.

\begin{figure}[htb]
\centering
\subfigure[$l=0$]{\label{fig:subfig:s12} 
\includegraphics[width=2.1in]{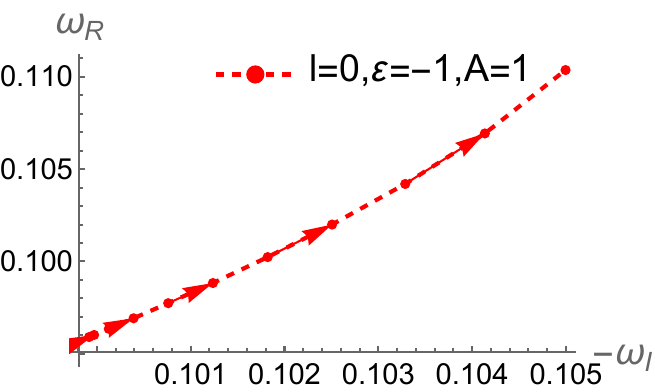}}
\hfill
\subfigure[$l=1$]{\label{fig:subfig:s22} 
\includegraphics[width=2.1in]{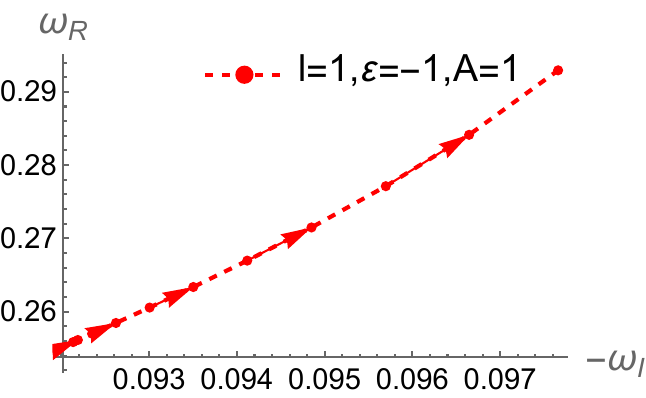}}
\hfill
\subfigure[$l=2$]{\label{fig:subfig:s32} 
\includegraphics[width=2.0in]{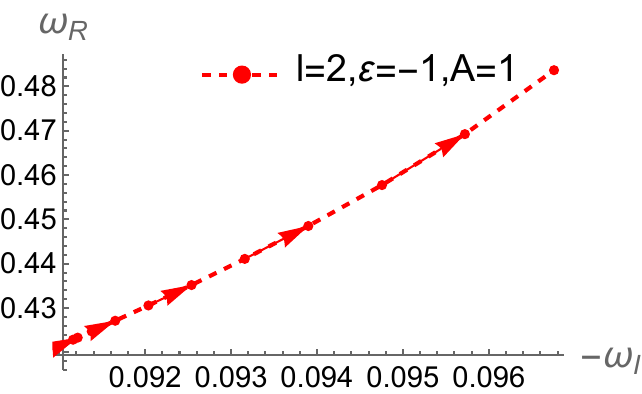}}
\hfill
\caption{The behaviors of real and imaginary parts of the \textbf{2nd} branch black hole solution's QNMs under \textbf{scalar perturbations} for different $l$ with $M=1, L=1, \lambda=1$. The arrows show the direction of the increase of black hole charge $Q$ from 0 to 1.}\label{figS2}
\end{figure}

\begin{figure}[htb]
\centering
\subfigure[$l=1$]{\label{fig:subfig:e1} 
\includegraphics[width=2.1in]{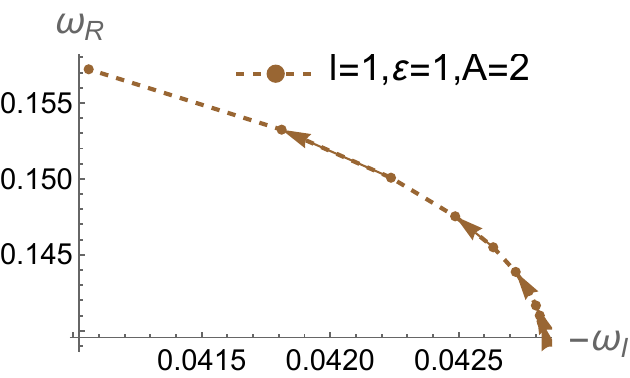}}
\hfill
\subfigure[$l=2$]{\label{fig:subfig:e2} 
\includegraphics[width=2.1in]{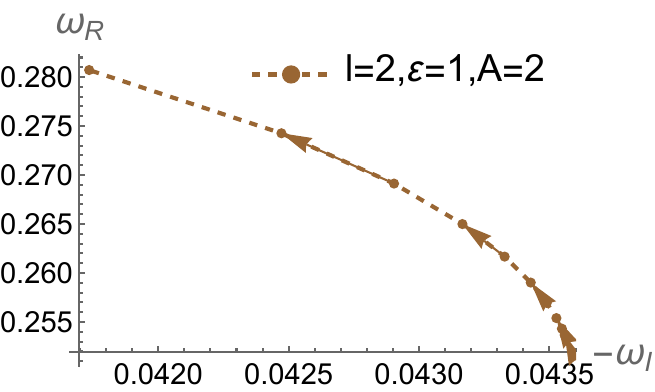}}
\hfill
\subfigure[$l=3$]{\label{fig:subfig:e3} 
\includegraphics[width=2.0in]{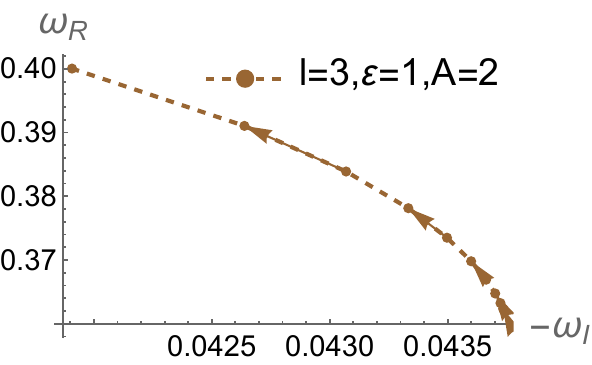}}
\hfill
\subfigure[$l=1$]{\label{fig:subfig:e11} 
\includegraphics[width=2.1in]{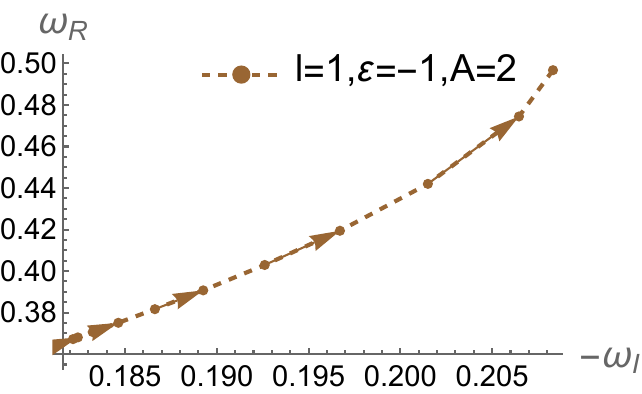}}
\hfill
\subfigure[$l=2$]{\label{fig:subfig:e21} 
\includegraphics[width=2.1in]{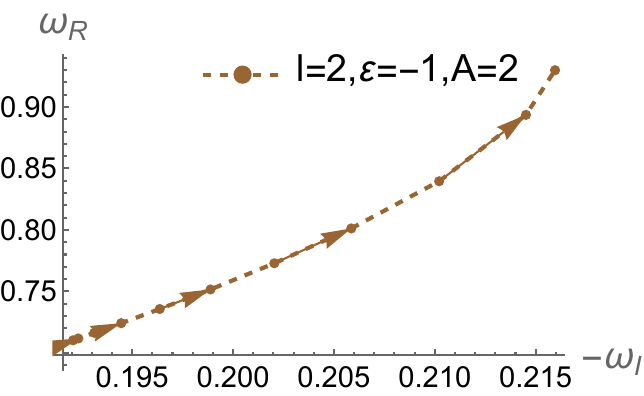}}
\hfill
\subfigure[$l=3$]{\label{fig:subfig:e31} 
\includegraphics[width=2.0in]{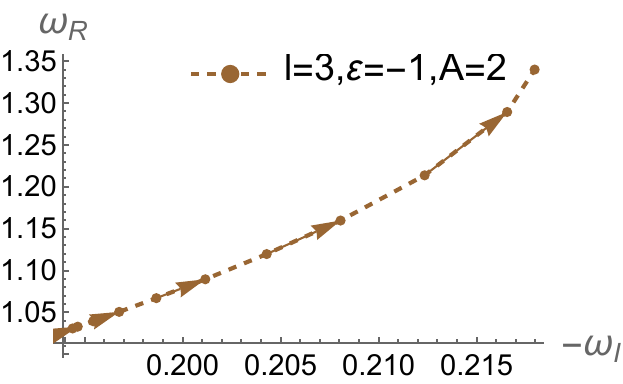}}
\hfill
\caption{The behaviors of real and imaginary parts of the \textbf{1st} branch black hole solution's QNMs under \textbf{Electromagnetic  perturbations} for different $l$ with  $M=1, L=1, \lambda=1$. The arrows show the direction of the increase of black hole charge $Q$ from 0 to 1.}\label{figE1}
\end{figure}

\begin{figure}[htb]
\centering
\subfigure[$l=1$]{\label{fig:subfig:e12} 
\includegraphics[width=2.1in]{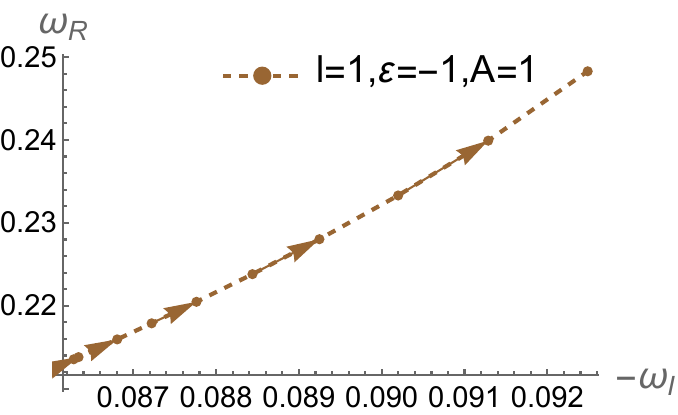}}
\hfill
\subfigure[$l=2$]{\label{fig:subfig:e22} 
\includegraphics[width=2.1in]{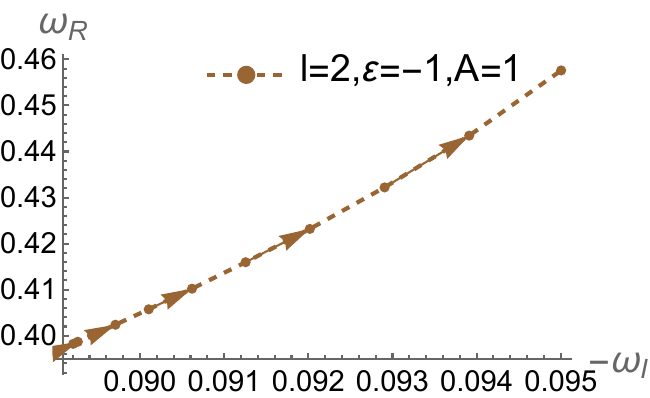}}
\hfill
\subfigure[$l=3$]{\label{fig:subfig:e32} 
\includegraphics[width=2.0in]{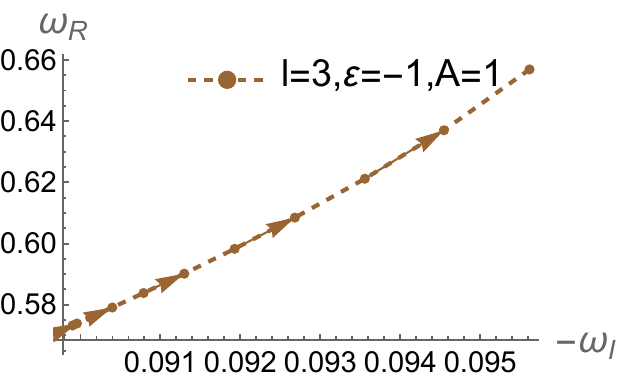}}
\hfill
\caption{The behaviors of real and imaginary parts of the \textbf{2nd} branch black hole solution's QNMs under \textbf{Electromagnetic perturbations} for different $l$ with $M=1, L=1, \lambda=1$. The arrows show the direction of the increase of black hole charge $Q$ from 0 to 1.}\label{figE2}
\end{figure}

\begin{figure}[htb]
\centering
\subfigure[$l=2$]{\label{fig:subfig:g1} 
\includegraphics[width=2.1in]{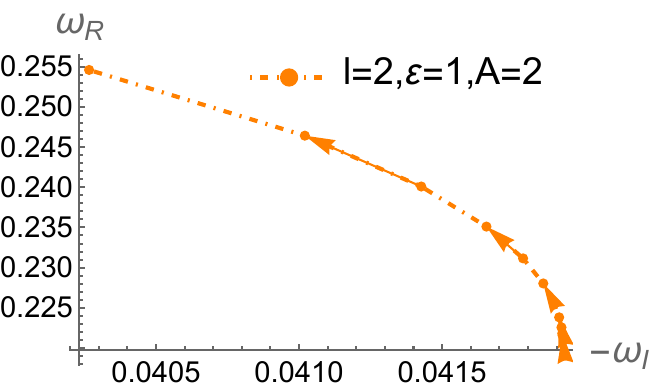}}
\hfill
\subfigure[$l=3$]{\label{fig:subfig:g2} 
\includegraphics[width=2.1in]{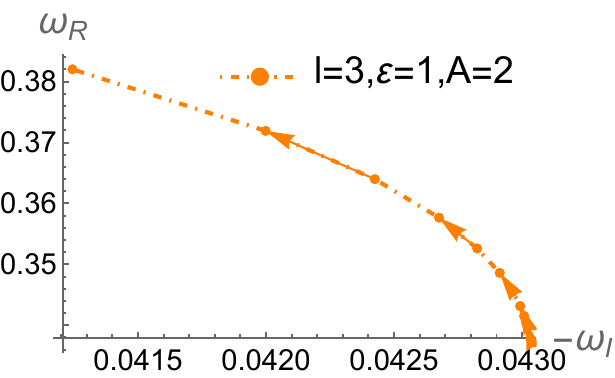}}
\hfill
\subfigure[$l=4$]{\label{fig:subfig:g3} 
\includegraphics[width=2.0in]{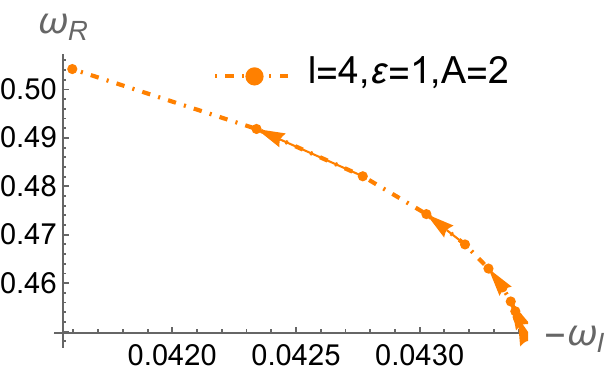}}
\hfill
\subfigure[$l=2$]{\label{fig:subfig:g11} 
\includegraphics[width=2.1in]{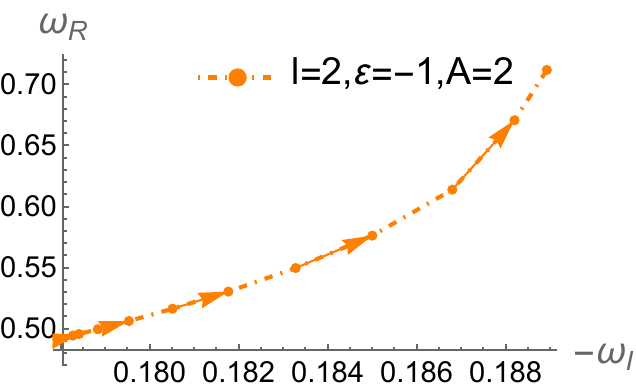}}
\hfill
\subfigure[$l=3$]{\label{fig:subfig:g21} 
\includegraphics[width=2.1in]{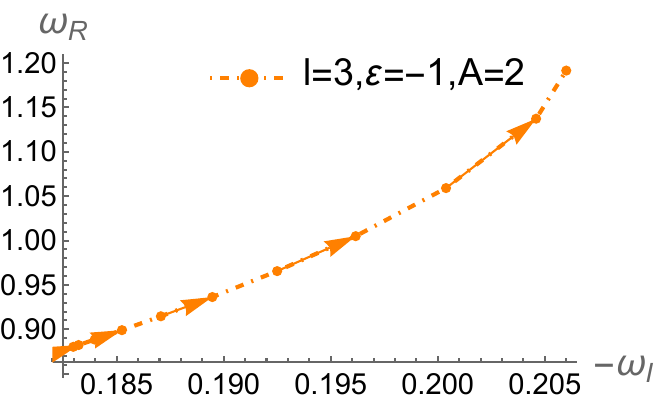}}
\hfill
\subfigure[$l=4$]{\label{fig:subfig:g31} 
\includegraphics[width=2.0in]{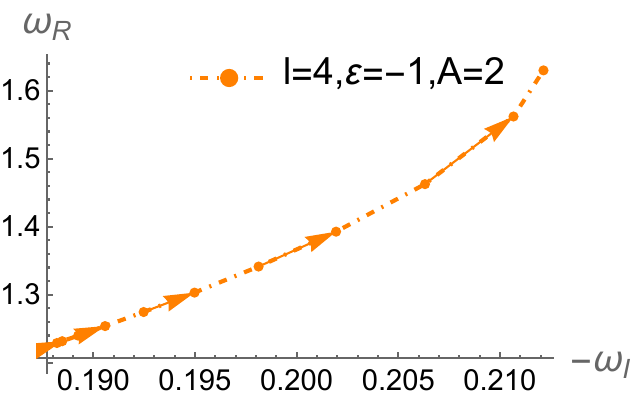}}
\hfill
\caption{The behaviors of real and imaginary parts of the \textbf{1st} branch black hole solution's QNMs under \textbf{Gravitational perturbations} for different $l$ with $M=1, L=1, \lambda=1$. The arrows show the direction of the increase of black hole charge $Q$ from 0 to 1.}\label{figG1}
\end{figure}

\begin{figure}[htb]
\centering
\subfigure[$l=2$]{\label{fig:subfig:g12} 
\includegraphics[width=2.1in]{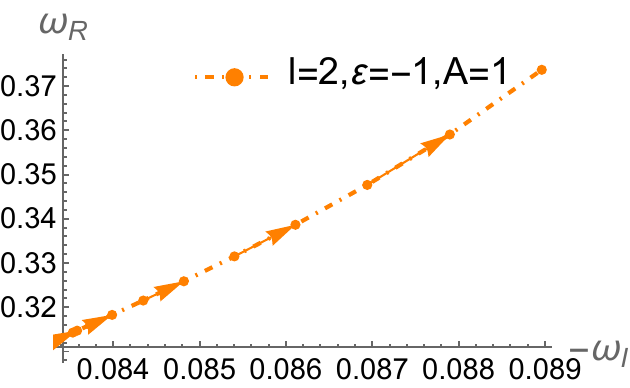}}
\hfill
\subfigure[$l=3$]{\label{fig:subfig:g22} 
\includegraphics[width=2.1in]{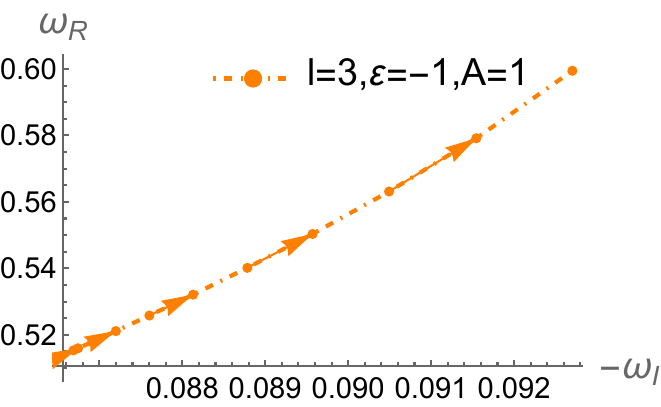}}
\hfill
\subfigure[$l=4$]{\label{fig:subfig:g32} 
\includegraphics[width=2.0in]{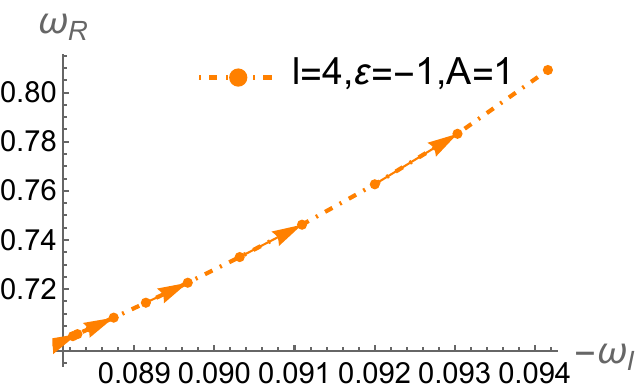}}
\hfill
\caption{The behaviors of real and imaginary parts of the \textbf{2nd} branch black hole solution's QNMs under \textbf{Gravitational perturbations} for different $l$ with $M=1, L=1, \lambda=1$. The arrows show the direction of the increase of black hole charge $Q$ from 0 to 1.}\label{figG2}
\end{figure}

We have discussed in detail the QNM behaviors of scalar perturbations in the 1st branch black hole solution. Similar results are found for the second branch solutions as well as for \textbf{electromagnetic and gravitational perturbations}, which are exhibited in Figs.\ref{figS2}-\ref{figG2}. The behavior of their QNMs aligns closely with the analysis presented earlier. Therefore, we do not repeat the detailed explanation here for these cases, as their characteristics exhibit similar trends across the different types of perturbations.

\section{Closing Remarks}
\label{sec6}

Considering black holes surrounded by a fluid of strings in four-dimensional Rastall
gravity, we discussed the scalar, electromagnetic,
and gravitational perturbations on these black holes. We applied the AIM to numerically compute the three perturbations and investigate the influence of the black hole
charge $Q$, angular momentum $l$ and the model parameters $\varepsilon$ on the quasinormal frequencies.

Specifically, for scalar perturbations and $\epsilon = 1$, we found that the absolute value of the imaginary part $|\omega_I|$ decreases with increasing $Q$, indicating that the perturbations decay more slowly as the charge grows. Conversely, for $\epsilon = -1$, $|\omega_I|$ increases with $Q$, suggesting that the perturbations dissipate more quickly. 
For the real part of the quasinormal frequency $|\omega_R|$, we observed a general trend of increasing frequency with both increasing black hole charge $Q$ and angular momentum $l$. This indicates that the perturbations gain energy as the black hole becomes more charged or as the angular momentum of the perturbation increases, leading to higher oscillation frequencies. Importantly, these behaviors were consistent across both branches of black hole solutions and for all types of perturbations studied—scalar, electromagnetic, and gravitational fields.

Overall, our study emphasizes the intricate relationship between black hole parameters and the behavior of QNMs, offering insights into the stability and dynamics of black holes in the context of Rastall gravity. Future research could further explore the implications of these results for holographic dualities and the broader implications for black hole thermodynamics.

 \vspace{1cm}
{\bf Acknowledgments}
 \vspace{1cm}

M. Z. is supported by Natural Science Basic Research Program of Shaanxi (Program No.2023-JC-QN-0053). D. C. Z is supported by the Natural
Science Foundation of China (NSFC) (Grant No.12365009) and Natural Science Foundation of Jiangxi Province (Grant No. 20232BAB201039).

\end{document}